# A Bayesian Approach to Calibrating High-Throughput Virtual Screening Results and Application to Organic Photovoltaic Materials

E.O. Pyzer-Knapp[a], G.N. Simm[b], and A. Aspuru Guzik[a] *

A novel approach for calibrating quantum-chemical properties determined as part of a high-throughput virtual screen to experimental analogs is presented. Information on the molecular graph is extracted through the use of extended connectivity fingerprints, and exploited using a Gaussian process to calibrate both electronic properties such as frontier orbital energies, and optical gaps and device properties such as short circuit current density, open circuit voltage and power conversion efficiency. The Bayesian nature of this process affords a value for uncertainty in addition to each calibrated value. This allows the researcher to gain intuition about the model as well as the ability to respect its bounds.

## Introduction

High-throughput virtual screening (HTVS) is a popular method for accelerating the discovery of both materials and pharmaceutical leads[1–17]. In HTVS, simplifications or approximations are often required to make this procedure computationally tractable, resulting in subtle differences between experimental and theoretical property definitions. In these cases, it is likely that a calibration will be performed on the calculated results, in order to facilitate comparison between calibrated and experimental data.

In this study, we investigate the relationship between calculated and experimentally observed values for electronic and device properties of organic photovoltaics. This area has been the subject of many high-throughput screening efforts[7–9], such as the Harvard Clean Energy Project (CEP)[6,18] in which errors in the model due to *in vacuo* calculations and oligomer vs polymer results were accounted for using an empirical linear calibration. Additionally, von Lilienfeld *et al*. have used a similar 'adjustment' method to predict the results of computationally expensive models from much simpler, cheaper, calculations[19].

Here we present an advance upon this calibration technique, which takes into account both quantum chemical information, and information about the molecular graph. In addition, this technique reports an uncertainty alongside each calibration – providing a confidence that the method is being used appropriately.

## Computational Methods and Theoretical Background

### Experimental Results and Theoretical Calculations

We recently reported the Harvard Organic Photovoltaic Dataset (HOPV15), which contains experimental results for 266 donor materials from bulk heterojunction devices, alongside corresponding quantum-chemical calculations performed using the BP86[20,21], B3LYP[20,22], M06-2x[23,24] and PBE0[25,26] functionals and the def2-SVP[27] basis set on the BP86/def2-SVP optimized geometry.

Within HOPV15, in order to simplify the conformational landscape, and only sample conformers whose electronic structure contributes to changes in photovoltaic efficiency, calculations are performed on the photovoltaic core – i.e. the molecule with any solubilizing long hydrocarbon chains replaced with a single methyl group. If this reduction resulted in two experimental results referring to the same 'pruned' molecule, the result set containing the larger experimental value for the power conversion efficiency (PCE) was used, since this best represents the potential of the core.

The highest occupied molecular orbital (HOMO), lowest unoccupied molecular orbital (LUMO), power conversion efficiency (PCE), open circuit voltage ($V_{OC}$), and short circuit current density ($J_{SC}$), are used in this study, with computational values for PCE, $V_{OC}$ and $J_{SC}$ generated using the Scharber model[28], and all computational properties reported as Boltzmann averages over conformers.

### Relating Molecular Structure to the Accuracy of Calculated Values

Many models, including density functionals[22,23], are constructed by fitting to values contained within a dataset. Since it is possible to reduce this fitting to a set of well-defined parameters, these are known as parametric methods. Parametric methods perform exceptionally well within the bounds of the model, but are ill defined when these bounds are broken.[29,30] A good example of this is the failure of density functionals trained using ground state structures to reproduce transition states accurately[31].

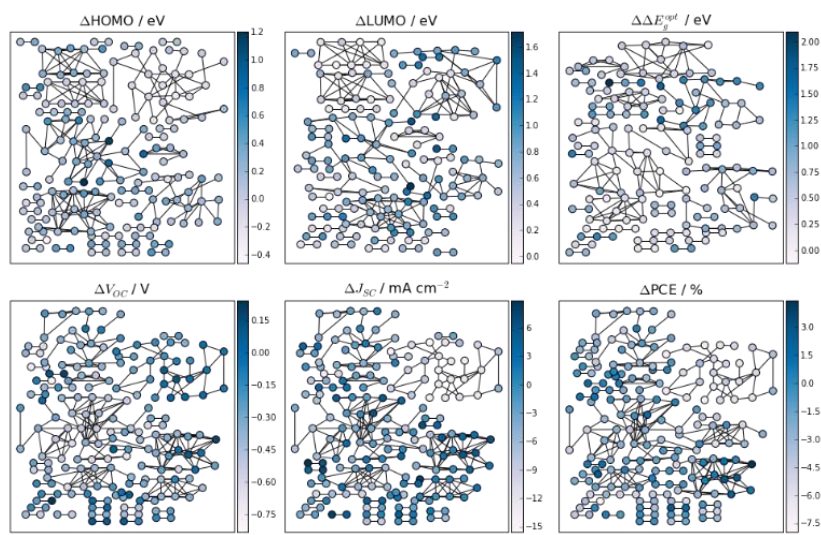

**Figure 1** Force directed graphs showing how the deviation between calculated and experimental values is cast into molecular space. The error in highest occupied molecular orbital energy, lowest unoccupied molecular orbital energy, optical gap (top row, left to right) and open circuit potential, short circuit current density and power conversion efficiency as calculated by the Scharber model (bottom row, left to right) are displayed via the colour of each node, which is itself representative of a molecule. Nodes are placed using the FDP algorithm, and connections are made between these nodes when their Tanimoto similarity between 512 bit, radius 2 Morgan circular fingerprints is > 0.65.

Additionally, potentially systematic sources of error are introduced through the difference between some calculated and experimental property definitions[28], the use of simplified model systems to reduce computational effort, and the use of an approximate density functional[32]. If systematic failings and errors are related to the chemical structures of the molecules in question[33], casting this problem into molecular space could afford a method for applying appropriate corrections which take into account the chemical makeup of the molecules in question.

To test this hypothesis we construct force-directed graphs for each property for which we have both experimental and simulated values. In order to cast this problem into chemical space, we extract information on each molecule as the 512-bit extended-connectivity (Morgan circular) fingerprint[34], with the connectivity radius calculated at the two-bond level using the implementation in the RDKit[35]. Each molecule is a node on this graph and nodes are connected when the Tanimoto similarity[36] is > 0.65. The FDP[37] algorithm, contained within the package Graphviz[38] was then used to structure the graph so that the edge-length was related to the similarity of molecules (i.e. the closer two connected nodes are, the more similar the molecules). Similar approaches have been used to perform 'materials cartography' – mapping chemical structure to properties in order to locate promising new materials[39]. Each node on the graph was coloured to represent the error in the calculation, as defined by the difference between calculated and experimental values.

As can be seen from **Figure 1**, broadly speaking clusters formed which appear on this graph share a similar deviation from experiment – as represented by the fill colour. This suggests that by mapping this problem onto molecular space, we will be able to perform a *per cluster* correction, and improve the accuracy of our calibration between calculated and experimental properties.

**Gaussian Processes**

Gaussian process regressions (GPs) have been extensively studied by the machine learning community and are known for their sophisticated and consistent theory combined with relative computational tractability[40,41]. They have been somewhat used within the scientific community to build structure-property relationship rules[42–45], although remain remarkably under-utilized, given their potential for strong predictive power.

While a probability distribution describes random variables which are scalars or vectors, a GP describes a distribution over functions. Within the framework of Bayesian inference, we can make predictions on an unknown data based upon input to target mappings described in a prior. This prior includes a covariance function – sometimes known as a kernel – which maps the covariance between function values. We build our covariance function upon the Tanimoto similarity $T(x_p, x_q)$[36], utilizing the popular squared exponential form for the kernel function itself:

$$k(x_p, x_q) = \sigma_f^2 exp\left(-\frac{1}{2l}(1 - T(x_p - x_q))\right) + \delta_{pq}\sigma_n^2 \quad (1)$$

Here, $\sigma_f^2$ is the signal variance, $l$ is a length scale, $\delta_{pq}$ is the noise variance, and $\sigma_n^2$ is a Kroneka delta, which takes the value 1 when $p=q$ else 0. The values for these hyperparameters were trained to optimize the log marginal likelihood using the L-BFGS algorithm[46]. In order to make a prediction, the Gaussian process places weights upon the functions of the prior distribution depending on how likely they are to model the target function. Thus the posterior distribution is sampled providing predictions (means), and also uncertainties for these predictions (standard deviations).

In contrast to the earlier discussed parametric methods, a Gaussian process is non-parametric and thus very few assumptions need to be made about the target function. Additionally, the accuracy of a non-parametric method will only increase with the size of the prior (i.e. more data).

## Results and Discussion

Gaussian processes were used to learn the deviation of computational results from their experimental analogues – i.e. to learn the function which calibrates one to the other.

In order to assess the performance of the calibration in a quantitative manner, we utilize two measures of error; the Pearson R coefficient, and a weighted RMSD. The Pearson R coefficient is a measure of the liner correlation between two variables, and is bounded at 0 (no correlation) and 1 (perfect correlation). In this study, increases in the Pearson R coefficient when the calibration has been applied are strongly indicative of an improvement of the performance. For the RMS Error, we include a weight to each point related to the uncertainty in the prediction returned by the Gaussian process. This weight is derived through normalizing the standard deviations of each prediction returned by the Gaussian process against the most certain point in the prediction. If these were not included, the measures would assume that each point is equally certain, thus removing a key piece of information from the scoring function In this way, we do not punish a poor calibration if it is also known to be highly uncertain. This was not done for the Pearson R coefficient, since bounding cases calculated on the distribution of points suggested that the un-weighted metric was representative.

We first examine the performance of calibrating the electronic properties of these molecules; the HOMO, LUMO and gap. A plot of predicted and experimental values for these properties is shown in **Figure 2**. This plot shows the results when calibrating the B3LYP[20,22] functional and def2-SVP basis set[27] – the results for the other functionals can be seen in the ESI, but are broadly similar. The hue of each point is related to the certainty of each prediction – the lighter the point, the more uncertain it is. Points were calibrated on a leave-one-out basis, in which the prior was formed using all points except the point being calibrated, with the process being repeated until all points had been calibrated.

It can be seen that whilst for these properties the quantum-chemical method performs reasonably, there is significant improvement when the calibration is applied. This is especially true for the calculation of the gap – a key property in the prediction of the performance of photovoltaic materials. This may be due in part to the correction of a systematic error in assuming that the gap can be adequately described by the difference in energy levels of the LUMO and HOMO. For this assumption to hold well, the ionization potential (IP) and electron affinity (EA) would have to be well described by these frontier orbitals, which is not necessarily the case in DFT[47]. The success of this calibration does show, however, that it is not necessary to calculate the EA and IP explicitly to rectify this error. This is of particular importance in the realm of high-throughput virtual screening, where an increase in the necessary number of calculations per molecule can swiftly accumulate, resulting in a significant decrease in the size of the libraries which can be screened.

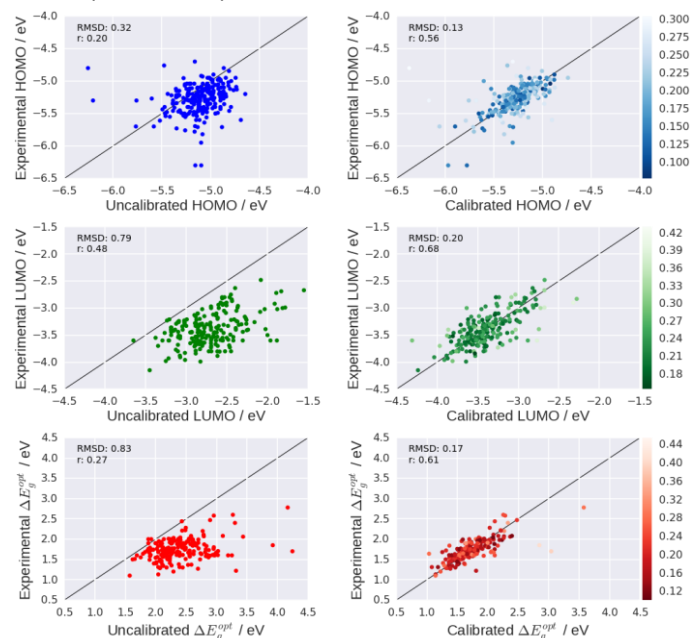

**Figure 2** The results of calibrating B3YP/def2-SVP quantum-chemical results for the Highest occupied molecular orbital (HOMO), lowest unoccupied molecular orbital (LUMO) and optical gap to the experimental HOPV15 data set. The uncertainty in the calibrated values is represented in the fill colour; the lighter the colour, the more uncertain the calibration.

Macroscopic properties, such as $J_{SC}$, $V_{OC}$, and PCE, present additional challenges to successful calibration. Since these properties have additional intermolecular contributions, capturing these in molecular fingerprints may prove challenging. Success in the prediction of lattice energy[48] and solubility[49] – properties both strongly related to intermolecular interactions – from the molecular structure does provide hope that these interactions can be somewhat captured, albeit in an implicit manner, in molecular fingerprints. Additionally, experimental measurements of device performance are notoriously noisy, introducing increasing amounts of uncertainty into the model. This method counters this through the application of some controlled noise to the data. The amount of noise was optimized against the log marginal likelihood to provide the function which was most robust to the data it was trained on.

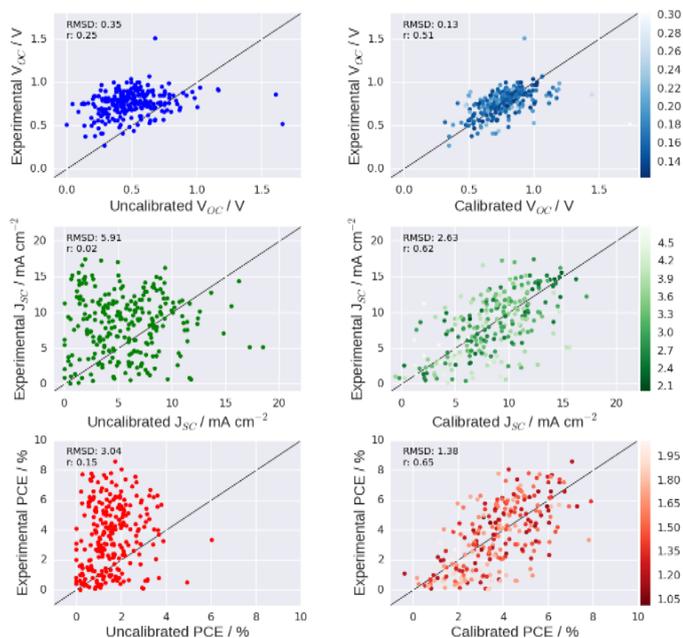

**Figure 3** The results of calibrating B3YP/def2-SVP quantum-chemical results for the open-circuit potential ($V_{OC}$), short circuit current density ($J_{SC}$), and power conversion efficiency (PCE) to the experimental HOPV15 data set. The uncertainty in the calibrated values is represented in the fill colour; the lighter the colour, the more uncertain the calibration.

**Figure 3** shows the results of the calibration of B3LYP[20,22]/def2-SVP[27] quantum chemical derived properties to the experimental $V_{OC}$, $J_{SC}$ and PCE extracted from HOPV15. These properties were extracted from the quantum-chemical values for the frontier orbitals using the Scharber model[28]. As with **Figure 2**, the colour of each calibrated point represents the uncertainty of the prediction, with light points being more uncertain.

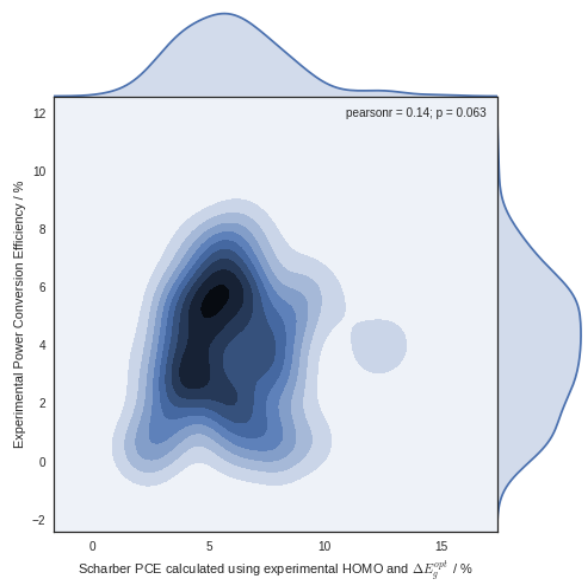

**Figure 4** The experimental power conversion efficiency shows a similar degree of correlation to the power conversion efficiency calculated using the Scharber model, but the low Pearson r value of 0.14 highlights the need for calibrating these values. This distributions plotted on the secondary axes show the distributions predicted by the Scharber model (X) experimentally observed (Y)

The calibrated results in **Figure 3** show significant improvements over the raw quantum-chemical results, demonstrating the power of this model. It can be seen that the PCE is particularly poorly predicted by the combination of the Scharber model and quantum-chemical results. The complementary study, in which the Scharber model is used in tandem with experimental values for the HOMO and gap shows a similar degree of correlation to that calculated with raw quantum-chemical results, and is shown in **Figure 4**. Since our calibrated results improve upon the use of the Scharber model with experimental values, we propose that errors arising from assumptions in the Scharber model seen in **Figure 4** are being implicitly corrected for in the Bayesian prior. We are currently investigating extending this methodology to provide such a method as a generic framework for building such data-driven, non-parametric models.

While the Pearson R values shown in **Figure 3** would suggest that the $V_{OC}$ is predicted the worst, it is the failure of the model to predict PCEs > ~ 5% which is particularly troubling; since these are the very values we are interested in. Since the Scharber model was built using experimental data[28], whilst the performance of the raw quantum-chemical inputs is not surprising, it strongly illustrated the importance of a model which provides a 'warning' when it is not being used in situations for which it is designed.

### Application to High Throughput Screening of Organic Photovoltaics

The low Pearson R and high weighted RMS error of the un-calibrated values highlight the importance of calibration when applying quantum-chemical results to certain photovoltaic properties.

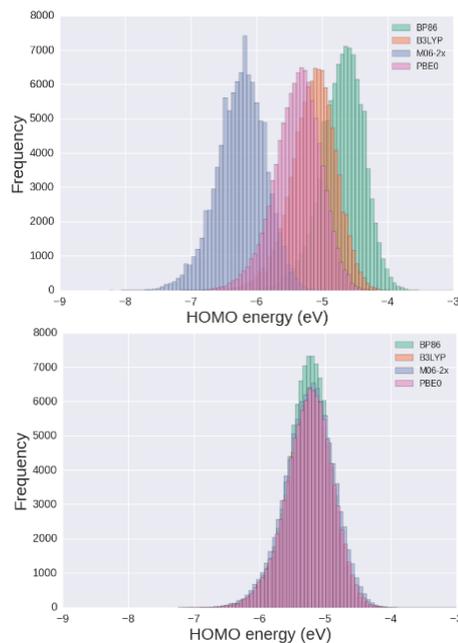

**Figure 5** Highest occupied molecular orbital (HOMO) energies (eV), Boltzmann averaged over conformers, as calculated by BP86, B3LYP PBE0 and M06-2x with the def2-SVP basis calculated for 100,000 molecules from the Clean Energy Project Database (top) and the values for the same set of molecules after calibration (bottom).

A good relationship between predicted and experimentally observed properties is paramount in the area of high-throughput screening, where design principles are built from trends in the data, and so errors and the re-ranking of candidates can have calamitous results In order to demonstrate this, 100,000 molecules were randomly selected from the CEPDB[50] to represent the results of a high-throughput virtual screen. The CEPDB contains the properties for each BP86[20,21]/def2-SVP[27] optimized geometry calculated at each of the functionals contained within this study. Since each functional is constructed in a different way, the properties calculated differ with the functional used.

**Figure 5** shows the distributions of HOMO energies Boltzmann averaged over each conformer in a molecule for the 100,000 molecule set. Values have been calculated using four different functionals: B3LYP[20,22], BP86[20,21], PBE0[25,26] and M06-2x[23,24] at the def2-SVP[27] basis. It can be clearly seen in the top plot (pre-calibration) that these values are very functional dependent.

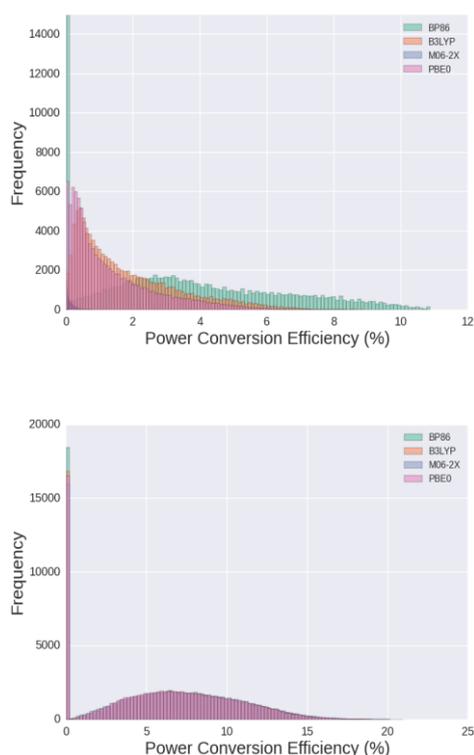

**Figure 7** The power conversion efficiency as calculated by BP86, B3LYP PBE0 and M06-2x with the def2-SVP basis pre (top) and post (bottom) calibration. Here, molecules with a predicted PCE less than 0 have been set to 0 to represent a physical result. The y-axis of the calculated PCE has been cut at 15,000 since the information is swamped by the highly skewed M06-2x distribution, which is shown alone in **Figure 6**. It can be seen that both the value and the distribution of values differs wildly between functionals. It can be seen that this functional dependence is removed once our calibration scheme is applied.

Analogous plots for the LUMO and gap, contained within the ESI, show the same dependence. Since these values are used to calculate the PCE and other photovoltaic properties, this is problematic as the properties, and identities, of candidates selected for further study should not depend upon the method used to calculate them. After calibration, however, these distributions strongly overlap, removing the functional dependence.

Thus, whichever functional is chosen for the study, a similar answer is returned – a behaviour much more congruent with experiment, and thus affording a greater confidence that the calculated property is representative of what would be observed if the experiment were to be performed. This argument extends into the PCE, which is the primary fitness function used for many high throughput virtual screening efforts. Since the PCE as calculated by the Scharber model is extremely sensitive to the energy levels of the HOMO and LUMO of the electron donor, small functional dependencies can manifest in vast differences in the distributions and values of PCEs obtained (**Figure 7**, top).

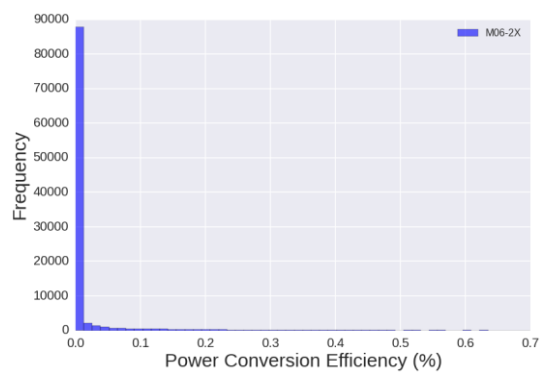

**Figure 6** The distribution of values for the Scharber power conversion efficiency calculated using the M06-2x functional at the def2-SVP basis are significantly skewed towards predicting a 0% power conversion efficiency. Here, all values for the power conversion efficiency predicted to be below 0 have been set to 0 to represent a physical result.

It can be seen that, whilst PBE0 and B3LYP display fairly similar values and distributions, M06-2X and BP86 do not share this desirable property. The fact that M06-2X predicts essentially all structures to have a PCE of 0 (**Figure 6**) is especially troubling and is most likely due to the overestimation of the gap by this functional which can been seen in the quantum-chemical predictions for this property by the MO6-2x functional contained within the ESI. If this functional alone had been selected, and no calibration performed, the result of the screening would have been incredibly pessimistic, and may have discouraged further research. Once our calibration scheme is applied (**Figure 7**, bottom), this functional dependence is removed; providing a greater confidence in the values returned.

## Conclusions

We show how a Bayesian approach to calibration, here implemented as a Gaussian process with a prior based upon relevant experimental observations, is a robust method for relating the results of quantum chemical calculations to experiment. For each of the properties in the HOPV15 dataset studied here - HOMO, LUMO, gap, PCE, $V_{OC}$, and $J_{SC}$ – values for the weighted RMS error between the calculated and experimental values, and the Pearson r correlation coefficient both improve significantly once the calibration is applied.

Additionally, we show how the application of this calibration method to the results of a high-throughput virtual screen removes the functional dependence of calculated properties, hence providing increased confidence that the result returned is directly comparable to experiment. This also increases the reliability and reproducibility of the candidate rankings, affording increased confidence that any extracted QSPR is an actual trend, and not an artefact of the choice of functional.

Finally, the Bayesian nature of our proposed calibration results in a confidence in each calibration point being returned. This is an invaluable tool, since it can inform the user that the scheme is being used for systems for which it is not designed, or for which the prior is not informative.

## Acknowledgements

A.A-G. and E.O.P-K acknowledges the Department of Energy through grant DE-SC0008733 for funding. This research would not have been possible without the use of the Harvard FAS Odyssey Cluster and support from FAS Research Computing. The authors also wish to thank IBM for organizing the World Community Grid and the WCG members for their computing time donations.

## Notes and references

1  E. O. Pyzer-Knapp, C. Suh, R. Gomez-Bombarelli, J. Aguilera-Iparraguirre and A. Aspuru-Guzik, *Annu. Rev. Mater. Res.*, 2015, **45**, null.
2  B. Huskinson, M. P. Marshak, C. Suh, S. Er, M. R. Gerhardt, C. J. Galvin, X. Chen, A. Aspuru-Guzik, R. G. Gordon and M. J. Aziz, *Nature*, 2014, **505**, 195–198.
3  B. K. Shoichet, *Nature*, 2004, **432**, 862–865.
4  J. Bajorath, *Nat. Rev. Drug Discov.*, 2002, **1**, 882–894.
5  M. D. Halls and K. Tasaki, *J. Power Sources*, 2010, **195**, 1472–1478.
6  J. Hachmann, R. Olivares-Amaya, S. Atahan-Evrenk, C. Amador-Bedolla, R. S. Sanchez-Carrera, A. Gold-Parker, L. Vogt, A. M. Brockway and A. Aspuru-Guzik, *J Phys Chem Lett*, 2011, **2**, 2241–2251.
7  I. Y. Kanal, S. G. Owens, J. S. Bechtel and G. R. Hutchison, *J. Phys. Chem. Lett.*, 2013, **4**, 1613–1623.
8  N. M. O'Boyle, C. M. Campbell and G. R. Hutchison, *J. Phys. Chem. C*, 2011, **115**, 16200–16210.
9  Y. Shu and B. G. Levine, *J. Chem. Phys.*, 2015, **142**, 104104.
10  Y. J. Colón, D. Fairen-Jimenez, C. E. Wilmer and R. Q. Snurr, *J. Phys. Chem. C*, 2014, **118**, 5383–5389.
11  M. D. Halls, P. J. Djurovich, D. J. Giesen, A. Goldberg, J. Sommer, Eric McAnally and M. E. Thompson, *New J. Phys.*, 2013, **15**, 105029.
12  M. D. Halls, D. J. Giesen, T. F. Hughes, A. Goldberg and Y. Cao, 2013, vol. 8829, pp. 882926–882926–6.
13  E. S. Kadantsev, P. G. Boyd, T. D. Daff and T. K. Woo, *J. Phys. Chem. Lett.*, 2013, **4**, 3056–3061.
14  S. Curtarolo, G. L. W. Hart, M. B. Nardelli, N. Mingo, S. Sanvito and O. Levy, *Nat. Mater.*, 2013, **12**, 191–201.
15  M. Korth, *Phys. Chem. Chem. Phys.*, 2014, **16**, 7919–7926.
16  C. E. Wilmer, M. Leaf, C. Y. Lee, O. K. Farha, B. G. Hauser, J. T. Hupp and R. Q. Snurr, *Nat. Chem.*, 2012, **4**, 83–89.
17  A. Jain, S. P. Ong, G. Hautier, W. Chen, W. D. Richards, S. Dacek, S. Cholia, D. Gunter, D. Skinner, G. Ceder and K. A. Persson, *APL Mater.*, 2013, **1**, 011002.
18  J. Hachmann, R. Olivares-Amaya, A. Jinich, A. L. Appleton, M. A. Blood-Forsythe, L. R. Seress, C. Roman-Salgado, K. Trepte, S. Atahan-Evrenk, S. Er, S. Shrestha, R. Mondal, A. Sokolov, Z. Bao and A. Aspuru-Guzik, *Energy Env. Sci*, 2014, **7**, 698.
19  R. Ramakrishnan, P. O. Dral, M. Rupp and O. A. von Lilienfeld, *J. Chem. Theory Comput.*, 2015, **11**, 2087–2096.
20  A. D. Becke, *Phys Rev A*, 1988, **38**, 3098–3100.
21  J. P. Perdew, *Phys Rev B*, 1986, **33**, 8822–8824.
22  A. D. Becke, *J. Chem. Phys.*, 1993, **98**, 5648–5652.
23  Y. Zhao and D. Truhlar, *Theor. Chem. Acc.*, 2008, **120**, 215–241.
24  Y. Zhao and D. G. Truhlar, *J. Chem. Theory Comput.*, 2007, **3**, 289–300.
25  J. P. Perdew, M. Ernzerhof and K. Burke, *J. Chem. Phys.*, 1996, **105**, 9982–9985.
26  J. P. Perdew, K. Burke and M. Ernzerhof, *Phys Rev Lett*, 1996, **77**, 3865–3868.
27  F. Weigend and R. Ahlrichs, *Phys Chem Chem Phys*, 2005, **7**, 3297–3305.
28  M. C. Scharber, D. Mühlbacher, M. Koppe, P. Denk, C. Waldauf, A. J. Heeger and C. J. Brabec, *Adv. Mater.*, 2006, **18**, 789–794.
29  P. J. Silva and M. J. Ramos, *Comput Theor Chem*, 2011, **966**, 120–126.
30  J.-W. Song, T. Tsuneda, T. Sato and K. Hirao, *Theor Chem Acc*, 2011, **130**, 851–857.
31  X. Xu, I. M. Alecu and D. G. Truhlar, *J. Chem. Theory Comput.*, 2011, **7**, 1667–1676.
32  A. J. Cohen, P. Mori-Sánchez and W. Yang, *Chem Rev*, 2012, **112**, 289–320.
33  D. E. Edwards, D. Y. Zubarev, A. Packard, W. A. Lester and M. Frenklach, *Phys. Rev. Lett.*, 2014, **112**, 253003.
34  D. Rogers and M. Hahn, *J. Chem. Inf. Model.*, 2010, **50**, 742–754.
35  G. Landrum, *RDKit: Open-source cheminformatics*, .
36  T. T. Tanimoto, *An elementary mathematical theory of classification and prediction, IBM Report (November, 1958), cited in: G. Salton, Automatic Information Organization and Retrieval*, McGraw-Hill New York, 1968.
37  T. M. J. Fruchterman and E. M. Reingold, *Softw Pr. Exper*, 1991, **21**, 1129–1164.
38  E. R. Gansner and S. C. North, *Softw - Pr. Exp*, 2000, **30**, 1203–1233.
39  O. Isayev, D. Fourches, E. N. Muratov, C. Oses, K. Rasch, A. Tropsha and S. Curtarolo, *Chem. Mater.*, 2015, **27**, 735–743.
40  M. Seeger, *Int J Neural Syst*, 2004, **14**, 69–106.
41  C. E. Rasmussen, *Gaussian processes for machine learning*, MIT Press, 2006.
42  O. Obrezanova, G. Csányi, J. M. R. Gola and M. D. Segall, *J. Chem. Inf. Model.*, 2007, **47**, 1847–1857.
43  F. R. Burden, *J. Chem. Inf. Comput. Sci.*, 2001, **41**, 830–835.
44  A. Schwaighofer, T. Schroeter, S. Mika, J. Laub, A. ter Laak, D. Sülzle, U. Ganzer, N. Heinrich and K.-R. Müller, *J. Chem. Inf. Model.*, 2007, **47**, 407–424.
45  P. Gao, A. Honkela, M. Rattray and N. D. Lawrence, *Bioinformatics*, 2008, **24**, i70–i75.
46  R. Byrd, P. Lu, J. Nocedal and C. Zhu, *SIAM J Sci Comput*, 1995, **16**, 1190–1208.
47  S. Kümmel and L. Kronik, *Rev. Mod. Phys.*, 2008, **80**, 3–60.

48  C. Ouvrard and J. B. O. Mitchell, *Acta Cryst Sect B*, 2003, **59**, 676–685.
49  J. L. McDonagh, N. Nath, L. D. Ferrari, T. van Mourik and J. B. O. Mitchell, *J. Chem. Inf. Model.*, 2014, **54**, 844–856.
50  *Harv. Clean Energy Proj. Database*.

# Supplimentary Information for : A Bayesian Approach to Calibrating High-Throughput Virtual Screening Results and Application to Organic Photovoltaic Materials

E.O. Pyzer-Knapp[a], G.N. Simm[b], and A. Aspuru Guzik[a] *

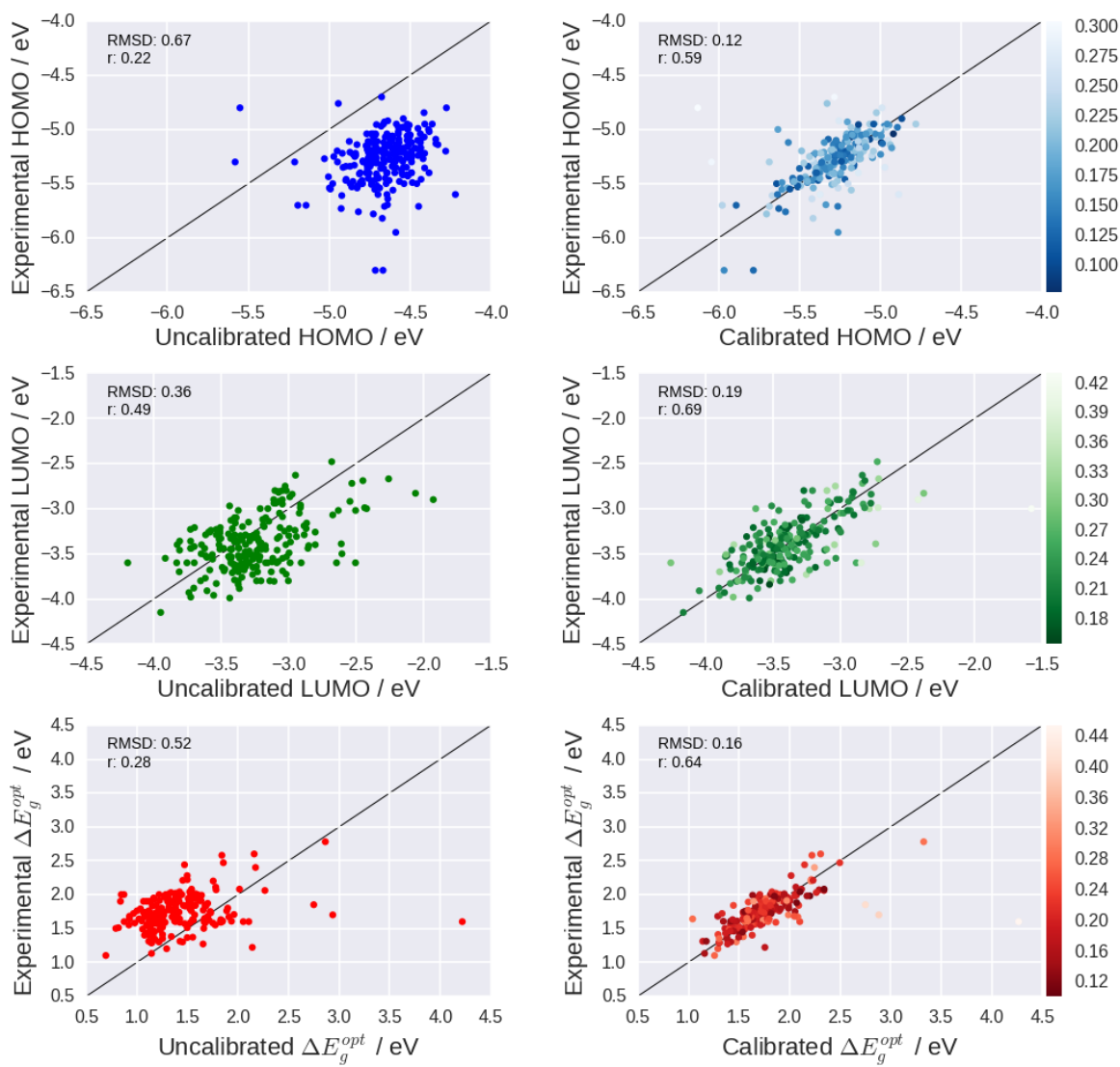

**Figure S1:** The results of calibrating BP86/def2-SVP quantum-chemical results for the Highest occupied molecular orbital (HOMO), lowest unoccupied molecular orbital (LUMO) and optical gap to the experimental HOPV15 data set. The uncertainty in the calibrated values is represented in the fill colour; the lighter the colour, the more uncertain the calibration.

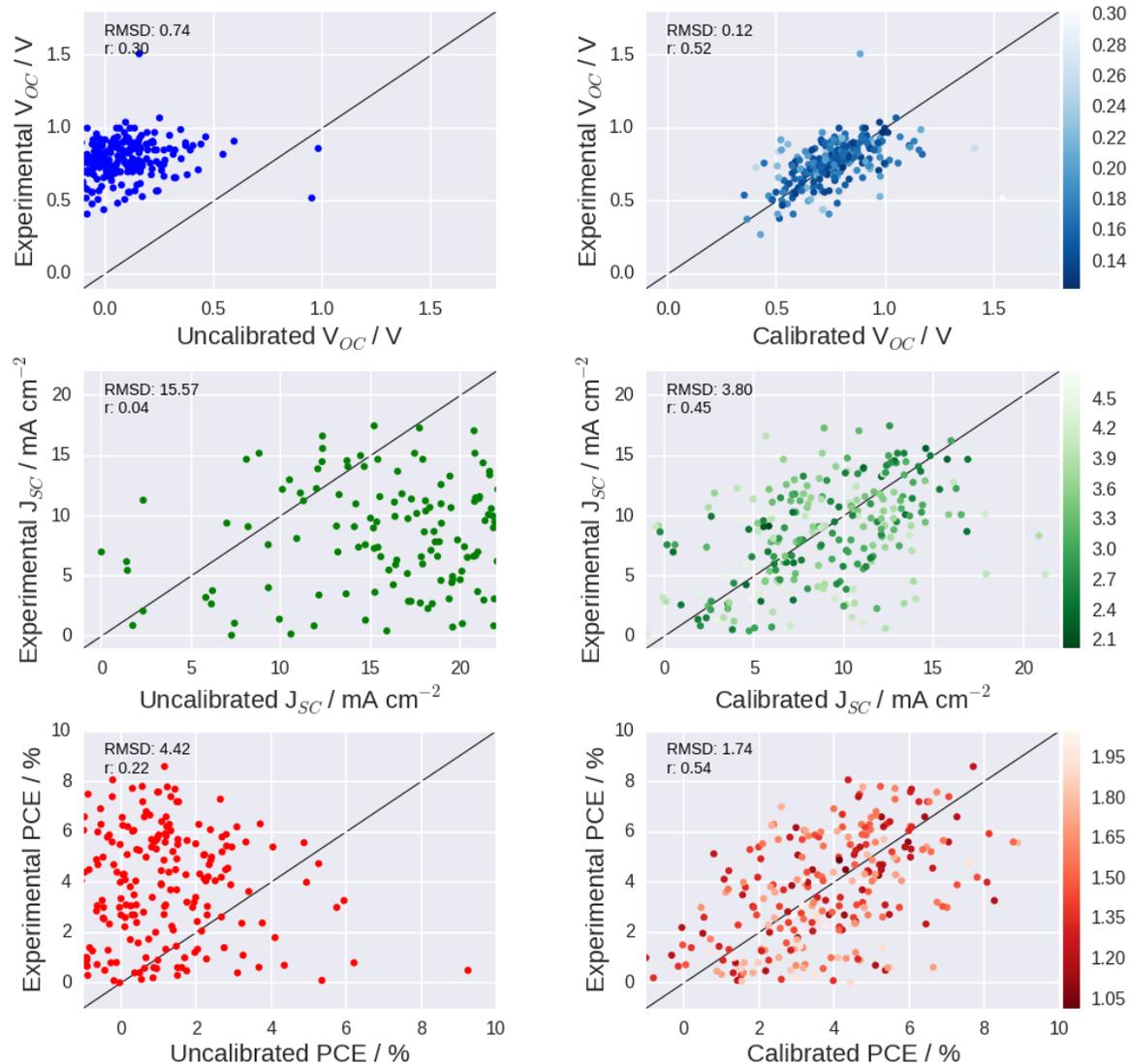

Figure S2: The results of calibrating BP86/def2-SVP quantum-chemical results for the open-circuit potential (VOC), short circuit current density (JSC), and power conversion efficiency (PCE) to the experimental HOPV15 data set. The uncertainty in the calibrated values is represented in the fill colour; the lighter the colour, the more uncertain the calibration.

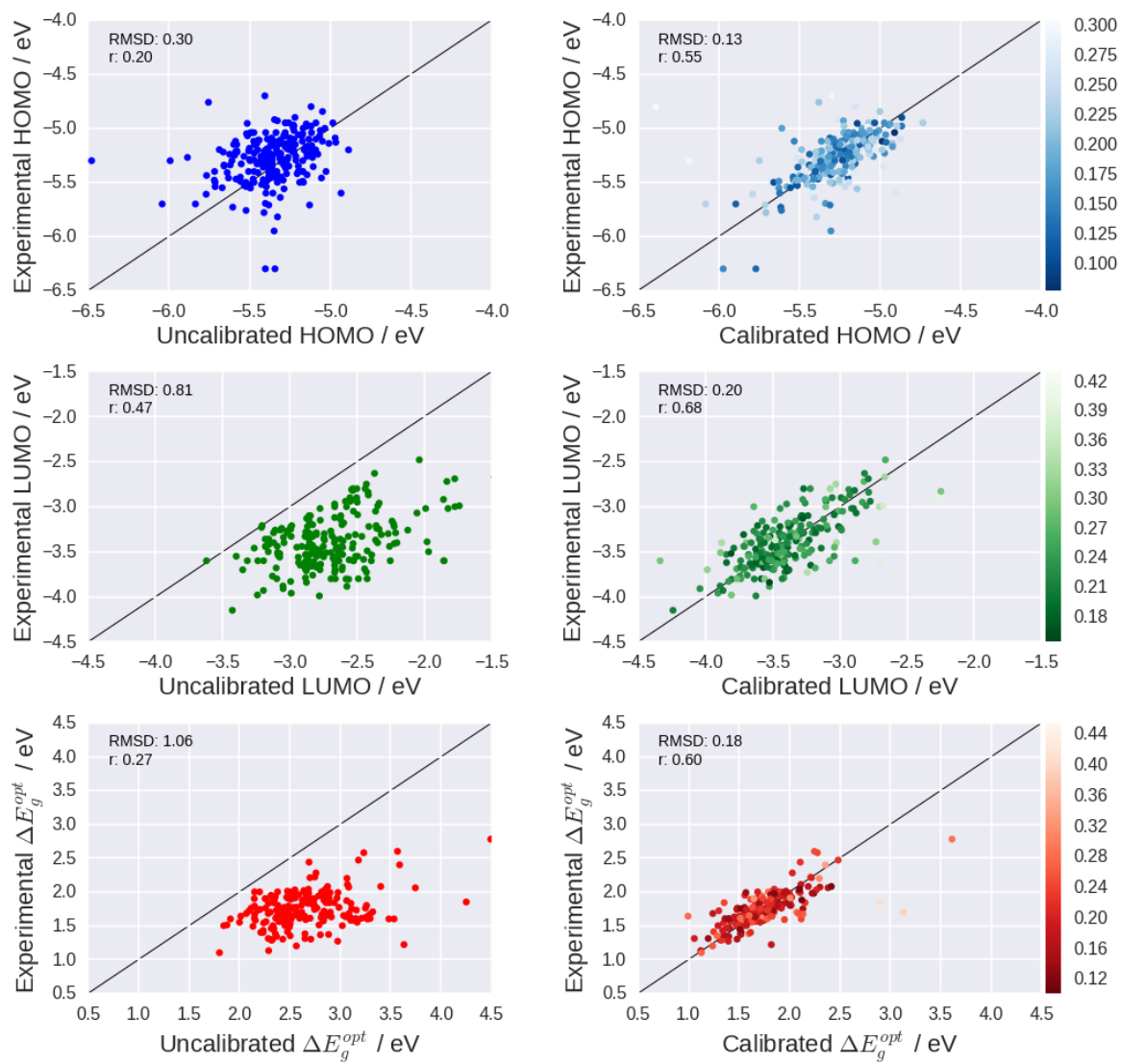

**Figure S3: The results of calibrating PBE0/def2-SVP quantum-chemical results for the Highest occupied molecular orbital (HOMO), lowest unoccupied molecular orbital (LUMO) and optical gap to the experimental HOPV15 data set. The uncertainty in the calibrated values is represented in the fill colour; the lighter the colour, the more uncertain the calibration.**

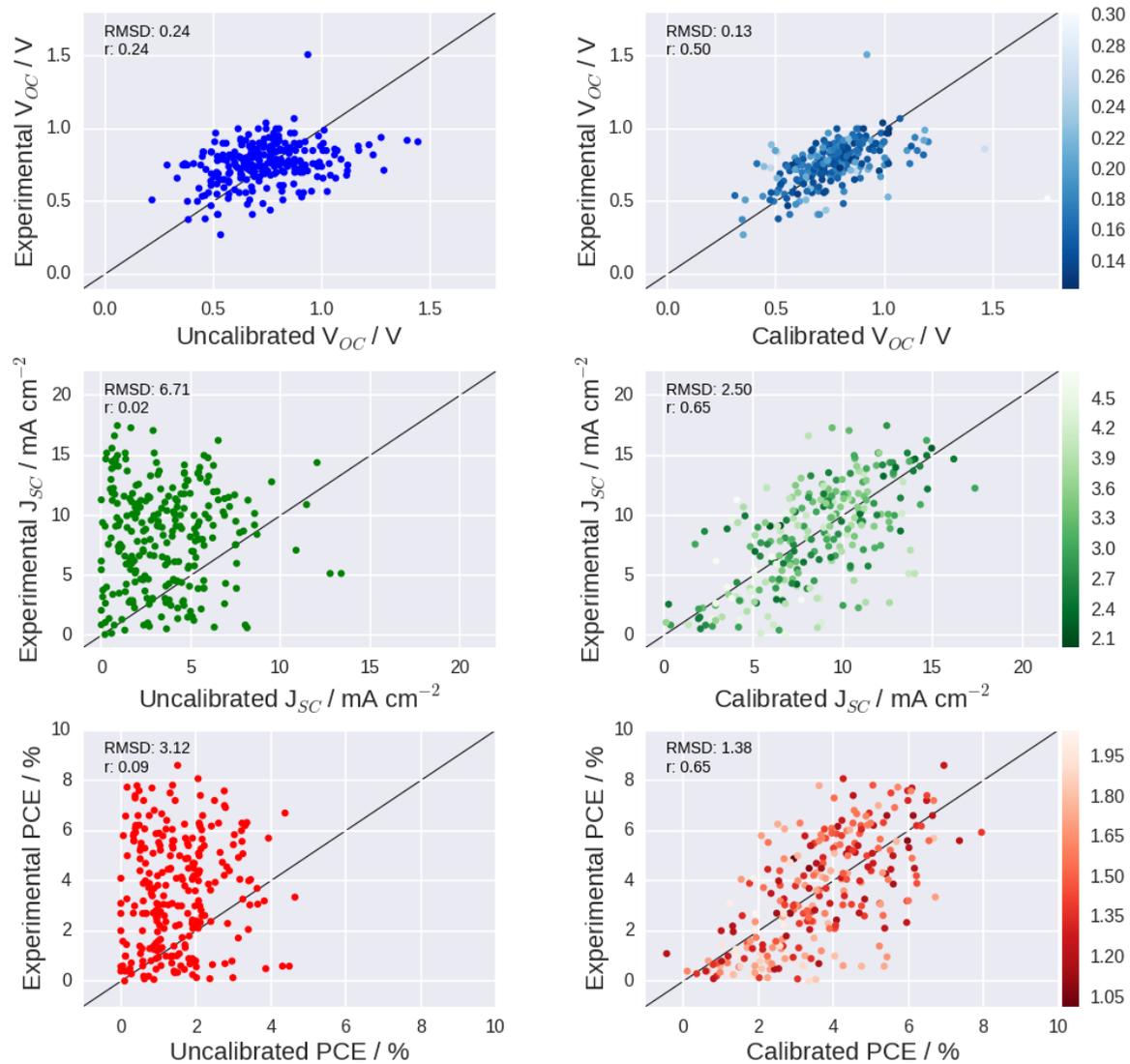

Figure S4: The results of calibrating PBE0/def2-SVP quantum-chemical results for the open-circuit potential (VOC), short circuit current density (JSC), and power conversion efficiency (PCE) to the experimental HOPV15 data set. The uncertainty in the calibrated values is represented in the fill colour; the lighter the colour, the more uncertain the calibration.

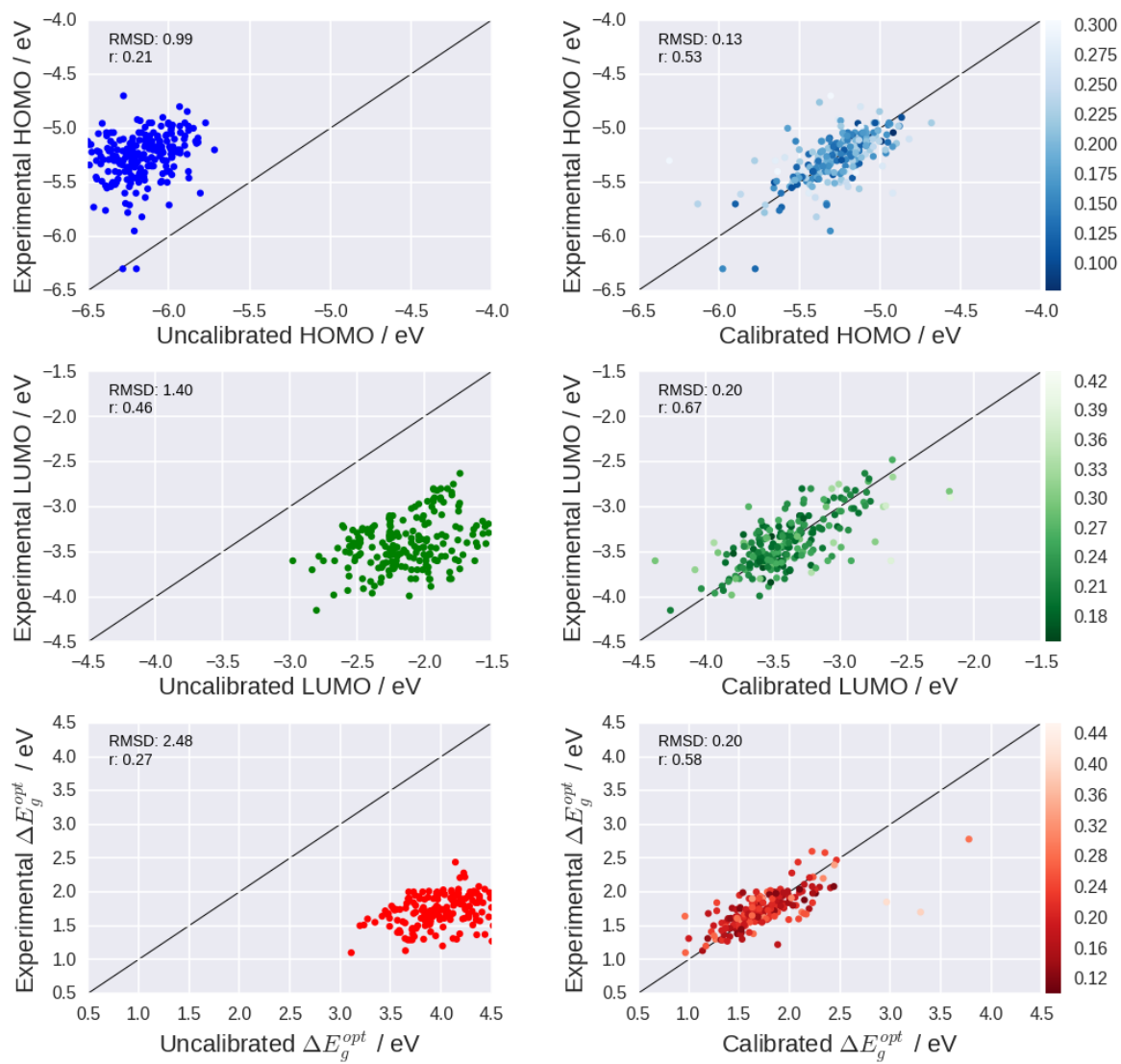

**Figure S5:** The results of calibrating M06-2X/def2-SVP quantum-chemical results for the Highest occupied molecular orbital (HOMO), lowest unoccupied molecular orbital (LUMO) and optical gap to the experimental HOPV15 data set. The uncertainty in the calibrated values is represented in the fill colour; the lighter the colour, the more uncertain the calibration.

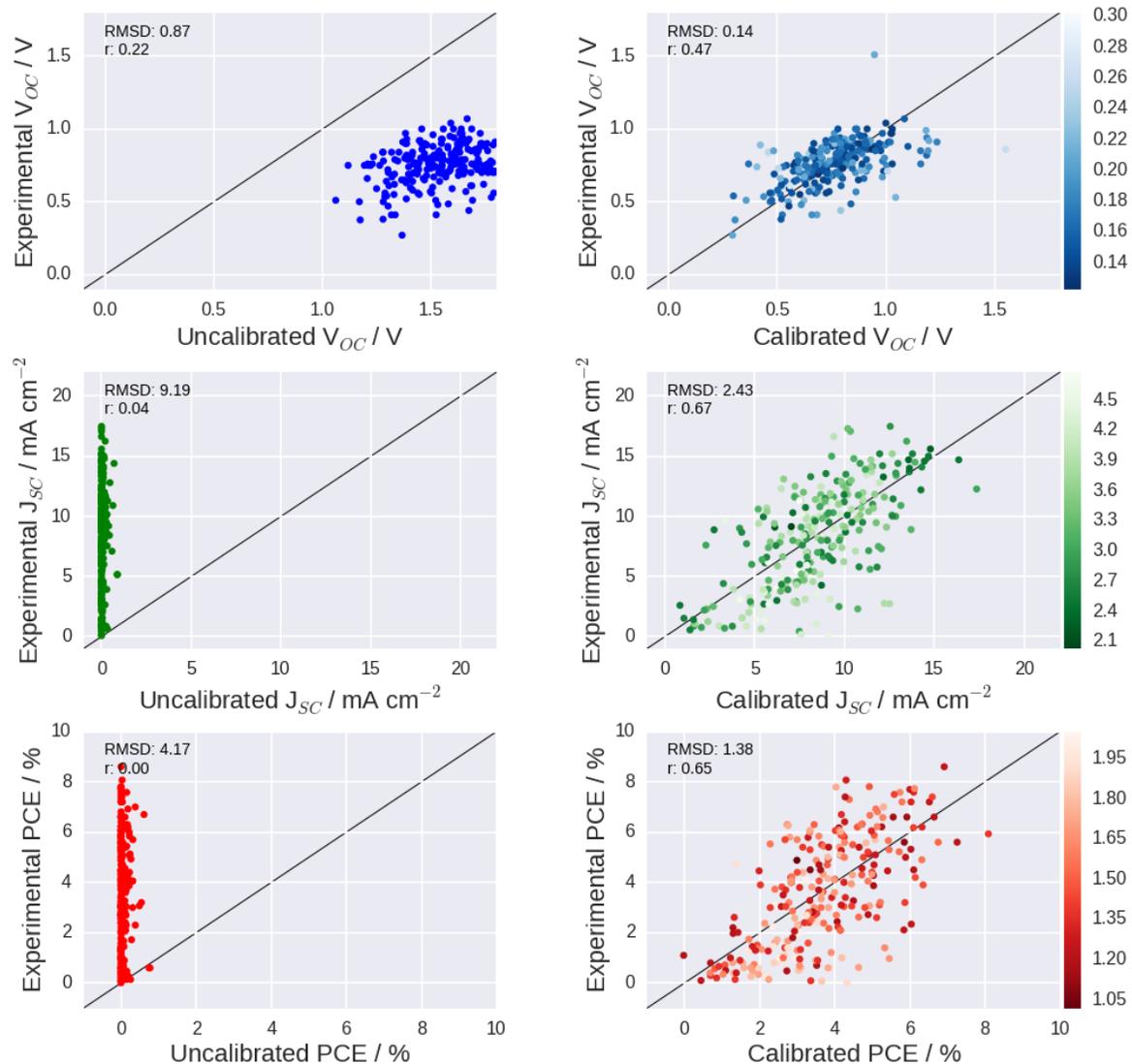

Figure S6: The results of calibrating M06-2X/def2-SVP quantum-chemical results for the open-circuit potential (VOC), short circuit current density (JSC), and power conversion efficiency (PCE) to the experimental HOPV15 data set. The uncertainty in the calibrated values is represented in the fill colour; the lighter the colour, the more uncertain the calibration.

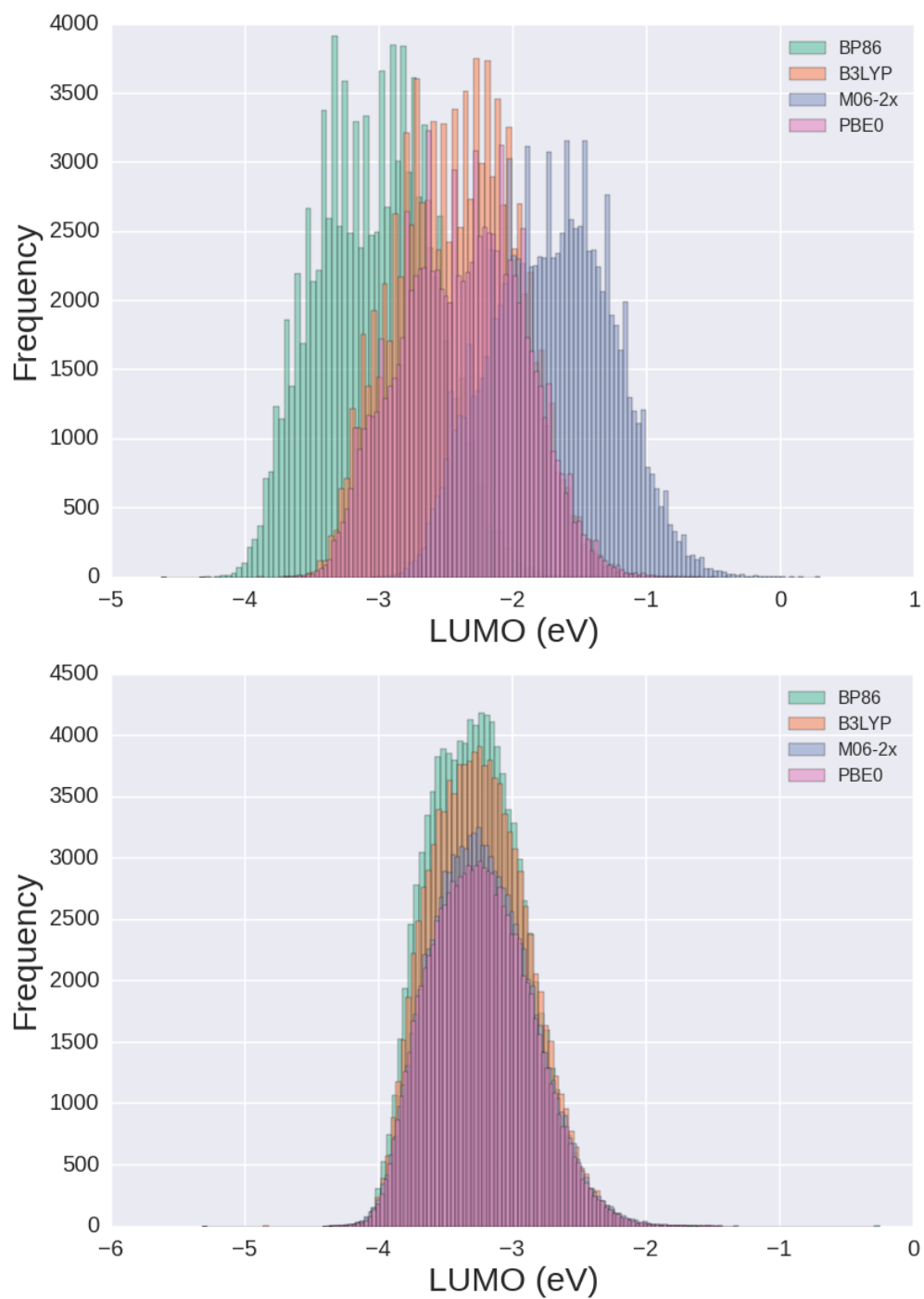

**Figure S7:** Lowest unoccupied molecular orbital (LUMO) energies (eV), Boltzmann averaged over conformers, calculated for 100,000 molecules from the Clean Energy Project Database (top) and the values for the same set of molecules after calibration (bottom).

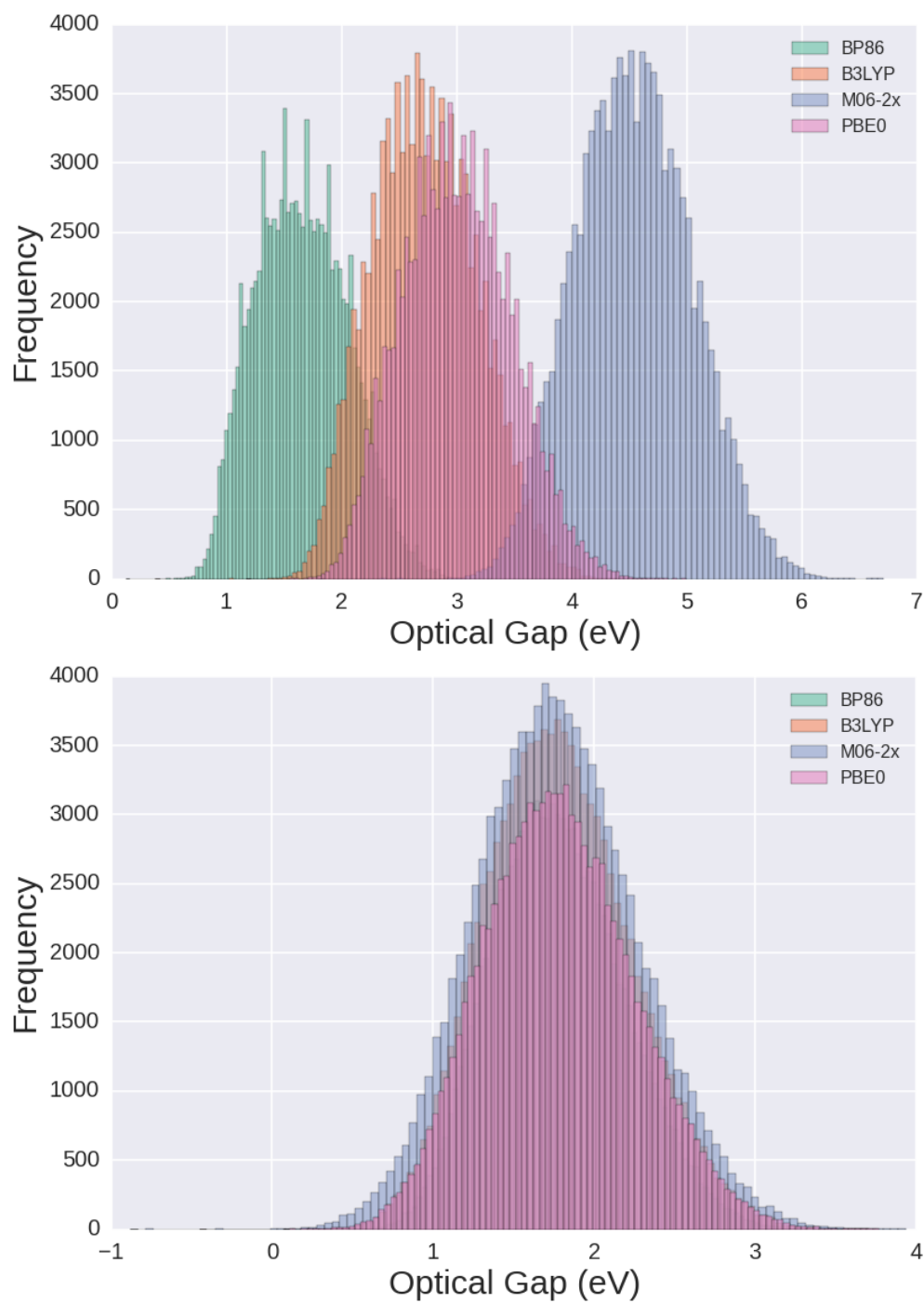

**Figure S8:** Optical gap energies (eV), Boltzmann averaged over conformers, calculated for 100,000 molecules from the Clean Energy Project Database (top) and the values for the same set of molecules after calibration (bottom).